\documentclass[fp,twocolumn]{jpsj3}
\usepackage{txfonts}

\title{
Apparent Ferroelectric Polarization Hysteresis and
a Simple Method to Observe Piezoelectric Strain Loops
with a Microphone
}

\author{
Mamoru \textsc{Fukunaga}\thanks{E-mail: fukunaga@ruri.waseda.jp}
}

\inst{
Independent, 
Okayama, 700-0081, Japan
}

\abst{
Extra components in series to non-ferroelectric capacitance can cause 
apparent ferroelectric $D$--$E$ hysteresis loops even with 
the double-wave method (DWM). 
Characteristics of fake loops are studied with simple circuit models, 
and suspicious loops of actual materials are found in some papers 
using the DWM. 
Inverse piezoelectric strain also reverses along with the polarization 
by the electric field, and $S$--$E$ loops are considered more reliable 
to prove ferroelectricity than the $D$--$E$ loops. 
But measurement of $S$--$E$ loops usually requires expensive instruments 
in contrast to $D$--$E$ loops with a simple circuit. 
A very simple method is developed to observe $S$--$E$ loops of bulk samples 
with an inexpensive small electret microphone 
and a little expansion to the circuit for $D$--$E$ loops. 
$S$--$E$ loops of a commercial ceramic capacitor by this method 
reveal that its $D$--$E$ loops apparent.
}

\begin{document}
\maketitle

\section{Introduction}

Ferroelectrics are materials which have the macroscopic polarization 
$P$ at the external electric field $E = 0$ and the remanent polarization 
$P\mathrm{_r}$ can be inverted by the $E$. 
Ferroelectricity is characterized by $P$--$E$ or $D$--$E$ hysteresis loops 
with $P$ or the electric flux density $D = P + \epsilon _0 E$. 
Principles of $D$--$E$ loop measurement are not changed 
from a traditional Sawyer-Tower circuit.\cite{DE0}
Applying an AC voltage $V = V\mathrm{_x}$ to a sample with electrodes, 
and the current $I$ through the sample is integrated 
to a standard capacitor of $C_0$. 
Voltage of the capacitor $V\mathrm{_y}$ gives the charge on the sample 
as $Q = C_0 V\mathrm{_y}$. 
$Q$--$V$ loops turn into $D$--$E$ loops with the relationship 
between $D = Q/A$ and $E = V\mathrm{_x}/d$, 
where $A$ is the area of the sample electrodes and $d$ is the thickness. 
$D$--$E$ loops of ferroelectrics exhibit hysteresis 
because $P$ reverses around the coercive field $E = \pm E\mathrm{_c}$.

When the sample is somewhat conductive, 
measured $D$ as the above increases and the $D$--$E$ loop 
inflates vertically, 
which looks like a hysteresis loop without ferroelectricity.\cite{DE1}
To remove such extra components in $D$, 
I have developed the double-wave method (DWM).\cite{DWM}
The method is similar to the Positive-Up-Negative-Down (PUND) method 
to evaluate ferroelectric thin films.\cite{PUND}
The DWM applies not continous bipolar waves but positive or negative 
half-waves twice. 
In each positive or negative two waves, 
the primary waves (called ``P'' and ``N'', and the pair as ``Pri'' in short) 
invert the polarization, 
while the secondary waves (``U'' and ``D'', the pair as ``Sec'') do not. 
Both the Pri and the Sec include other components 
except for the polarization, 
and the difference in $D$ between P and U, and between N and D (``Diff'') 
is considered as the ferroelectric polarization component only.

Extra components which increases the measured $D$ or $Q$ as the above 
are assumed to exist parallel to the ferroelectric component. 
In the case, $V$ is applied to each component equally 
and $Q$ is the sum of the components, therefore the DWM can subtract. 
In other words, the voltage on the sample capacitance 
always equals to the applied $V$, and its $Q$ is zero at $V = 0$. 
But if resistance exists in series to capacitance, 
voltage of the capacitance does not immediately become zero 
at $V = 0$ and the charge remains as apparent $P\mathrm{_r}$. 
This apparent $P\mathrm{_r}$ becomes different in P and U, 
and is not canceled by the DWM. 
Not only resistance but generally extra components 
series to capacitance can cause apparent hysteresis loops 
and the $P\mathrm{_r}$ by the DWM. 
Some other mechanisms to cause apparent loops have also been suggested.
\cite{FDET1,FDET2} 
As long as $D$ is measured as the current through the sample, 
apparent $D$--$E$ loops are unavoidable, 
and shapes of the loops are considered as the only key 
to distinguish true ferroelectric loops from fakes.

Ferroelectrics exhibit piezoelectric effects, 
and polarity of the effects can be inverted along with 
the remanent polarization. 
Polarization reversal by $E$ can also be observed as 
inverse piezoelectric strain $S$ versus $E$, 
and $S$--$E$ loops are considered more reliable than $D$--$E$ loops 
to confirm ferroelectricity. 
Piezoelectric coefficients are generally up to hundreds pC/N or pm/V, 
and measurement as the former unit, 
namely applying a stress and measuring induced charge is easy 
with an appropriate operational amplifier.\cite{QF} 
The latter, namely applying a voltage and measuring induced displacement 
including $S$--$E$ loops seems to require expensive instruments 
such as a laser micrometer.

I tried to observe the small displacement as a capacitance change, 
and found that sensitivity of an electret (condenser) microphone 
(EM or ECM) functioning with a similar principle is specified 
at a sound pressure of 1 Pa. 
Although EMs are small, inexpensive, and widely used as sound sensors, 
they respond to the change of $10^{-5}$ of ambient air pressure. 
The photoacoustic technique\cite{PA} seems to be using the sensitivity. 
I considered that the sensivity of an EM can be used 
to observe $S$--$E$ loops. 
This paper describes characteristics of apparent ferroelectric $D$--$E$ loops 
studied with circuit models, methods to observe $S$--$E$ loops with an EM, 
and $D$--$E$ and $S$--$E$ loops of a commercial lead zirconate titanate (PZT) 
ceramics and a commercial multilayer ceramic capacitor (MLCC) as examples.

\section{Experiments}

\begin{figure}
\centerline{\includegraphics[width=8.6cm]{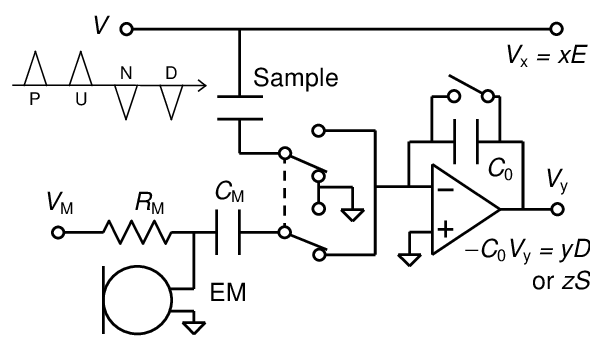}}
\caption{
A circuit for measuring $D$--$E$ and $S$--$E$ loops. 
$V\mathrm{_M}$ = 3.3 V, $R\mathrm{_M}$ = 1 k$\Omega$, 
$C\mathrm{_M}$ = 100 nF.
}
\label{fig1}
\end{figure}

Figure \ref{fig1} shows a circuit for measuring $D$--$E$ and $S$--$E$ loops. 
Output of an EM proportional to the pressure change 
and $S$ is measured as a charge through a capacitor $C\mathrm{_M}$, 
and a pair of two switches is used to choose $D$ and $S$ signals. 
(Although $C_0$ is also changed for $D$ or $S$, 
its switch is omitted in the figure.) 
$I$--$V$ curves are also measured instead of $D$--$E$ loops 
by replacing a standard capacitor $C_0$ with a resistor $R_0$. 
Voltage waveforms applied to the sample are generated 
and voltages proportional to $E$, $D$, and $S$ are measured 
by a homemade data aquisition board with a PC. 
The PC can also apply a continuous sine wave $V(t) = V_0 \sin\omega t$
and calculate average coefficients $Q\mathrm{_{1X}}$ and so on 
of a Fourier series written as 
$Q(t) = Q\mathrm{_{1X}} \sin\omega t - Q\mathrm{_{1Y}} \cos\omega t 
+ Q\mathrm{_{2X}} \sin 2\omega t - Q\mathrm{_{2Y}} \cos 2\omega t + \cdots $
like a lock-in amplifier. 
Waveforms used in the DWM are specified with a shape, 
an interval and a frequency. 
The shape is a triangular wave. 
The interval is defined as a time between the beginnings 
of the half-waves, which is typically 300 ms. 
The frequency equals to that of equivalent continuous waves, 
which is typically 20 Hz and the duration of each half-wave is 25 ms.

\begin{figure}
\centerline{\includegraphics[width=8.6cm]{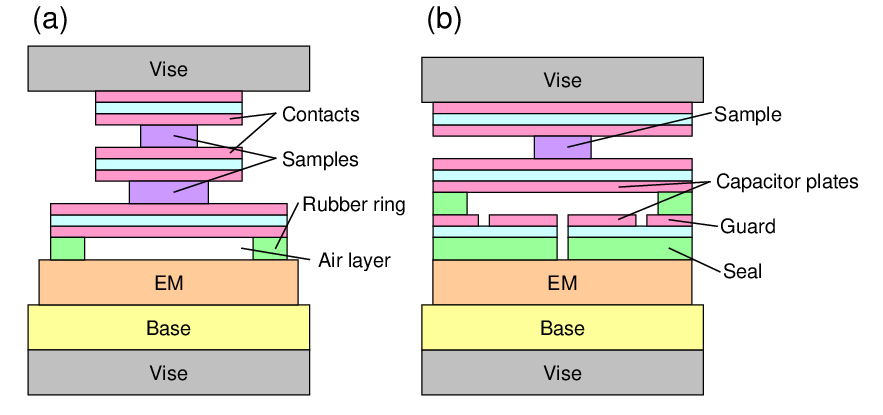}}
\caption{
(Color online) A cross section of a setup 
for measuring $S$--$E$ loops with an EM. 
The scale is vertically elongated except for an EM and a vise shrunk. 
(a) Two samples are stacked. 
(b) One sample is put and an air layer capacitor forms.
}
\label{fig2}
\end{figure}

Figure \ref{fig2}(a) shows a mechanical setup for $S$--$E$ loops. 
An EM is mounted on a base with the flat bottom. 
A rubber ring and the bottom contact plate on the EM 
makes an air layer. 
Samples and contact plates are also put on them, 
and a vise holds the whole stack. 
The contact plates are raw circuit boards with copper foil on both sides. 
Thin electric wires are attached to the contact plates. 
The rubber ring is cut from a rubber balloon to make it thin 
(approx. 0.2 mm). 
The vise is used to keep the total thickness constant, 
and is not tightened up. 
All parts except for the rubber ring are hard enough, 
and pressure of the air layer is expected to change proportionally 
to the piezoelectric strain of the sample, which is detected by the EM.

On calibration of $S$--$E$ loops, 
namely to convert EM signals into strain or displacement, 
Fig. \ref{fig2}(a) shows one method with two samples and 
one of them is used as a known standard. 
Figure \ref{fig2}(b) shows another method with one sample and 
an air layer capacitor. 
In this setup, a high AC voltage with the amplitude $V_0$ is applied 
to the air layer capacitor, and the charge is measured as $Q\rm{_{1X}}$. 
The capacitor has a guard plate on the bottom, 
and the charge excluding the guard $Q\rm{_X}$ and including $Q\rm{_I}$ 
are measured. 
From $V_0$, $Q\rm{_X}$, $Q\rm{_I}$ and the area of the bottom plate 
excluding the guard $A$, 
the distance $d$ and a peak of attraction $F\rm{_A}$ are calculated 
as $d = \epsilon _0 A V_0/Q\rm{_X}$ and 
$F\mathrm{_A} = Q\mathrm{_I} V_0/2d$. 
Little displacement by the attraction is calculated by $F\rm{_A}$ 
and elasticity of the rubber ring 
which is determined by the change of $d$ with weights instead of the vise.

PZT samples are broken pieces of a ceramic in a commercial 
piezoelectric sounder. 
Its dimensions are approximately 0.2 mm in thickness and 1.6 mm$^2$ in area. 
A commercial MLCC ($1.6 \times 0.8 \times 0.8\ \rm{mm}^3$) is also used. 
$S$--$E$ loops of the MLCC is measured with standing it 
to attach its external electrodes to contact plates. 
Internal electrodes in the MLCC are perpendicular 
to external electrode faces, 
and the sign of observed $S$ inverts.

\section{Results and discussion}

\subsection{DWM $D$--$E$ loops of a PZT and an MLCC}

\begin{figure}
\centerline{\includegraphics[width=8.6cm]{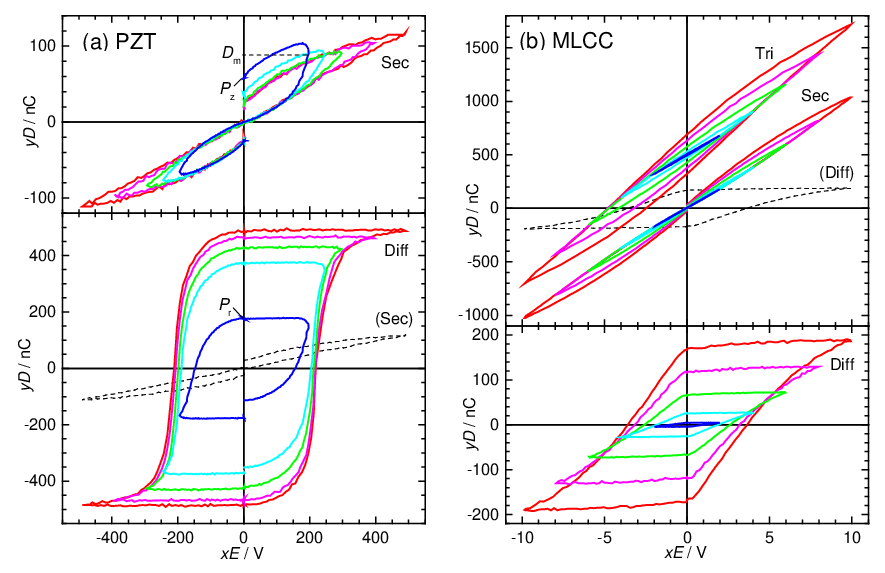}}
\caption{
(Color online) $D$--$E$ loops of (a) the PZT and (b) the MLCC samples 
with various amplitude $V_0$. 
Vertical scales for Diff and Sec groups are different.
}
\label{fig3}
\end{figure}

At first, Fig. \ref{fig3}(a) shows true ferroelectric $D$--$E$ loops 
of a PZT sample by the DWM. 
In the figures, although $y$ is area and $x$ is thickness 
for a plate-like sample, 
$yD$ and $xE$ means that $D$--$E$ loops are drawn 
as proportional raw $Q$--$V$ loops. 
As representative values of $D$ in Sec loops, 
$D\rm{_m}$ and $P\rm{_z}$ are defined as shown in the figure. 
$D = D\rm{_m}$ at the middle point of the half-loop 
and the maximum of $|E| = E_0$. 
$D\rm{_m}$ is proportional to permittivity as 
$D\mathrm{_m} = \epsilon_0 \epsilon\mathrm{_r} E_0$ 
for paraelectric materials. 
$D = P\rm{_z}$ at the end of the half-loop and $E = 0$. 
$P\rm{_z}$ is proportional to the conductivity and $D\rm{_m}$ includes 
$P\rm{_z}/2$ when $P\rm{_z}$ arises from the conductivity only. 
In this case, $P\rm{_z}$ does not mean the conductivity 
because it is not proportional to the amplitude $E_0$. 
Here $P\rm{_z}$ means a short-life polarization reversed by the $E$ 
but returns in the period of DWM interval. 
The loops were measured with various $V_0$, 
and it was considered that 
Diff loops by the DWM indicates reversible $P\rm{_r}$ for each $V_0$. 
Diff loops for small $V_0$, however, are not closed 
and $P\rm{_z}$ in $E > 0$ are larger than those for higher $V_0$. 
This result indicates that 
a part of polarization does not invert in Pri loops 
and inverts in Sec loops with small $V_0$, 
and in such a case $P\rm{_r}$ is underestimated. 
Asymmetric loops for positive and negative $E$ means 
the polarization reversal is easy for $E < 0$ in this case 
because irreversible polarization domain exists.

Figure \ref{fig3}(b) shows $D$--$E$ loops of an MLCC. 
Tri loops are measured by continuous triangular waves (not the DWM). 
Although temperature dependence is quite different, 
high capacitance MLCCs are made from barium titanate based ceramics, 
and they may have ferroelectricity. 
Diff loops clearly exhibit hysteresis. 
But $D$ changes monotonically from $E = 0$. 
$P\rm{_r}$ is not saturated and becomes larger as $V_0$ rises. 
These loops seem strange as ferroelectric hysteresis loops 
in comparison with those of PZT.

\subsection{Apparent ferroelectric $D$--$E$ loops of circuit models}

\begin{figure}
\centerline{\includegraphics[width=8.6cm]{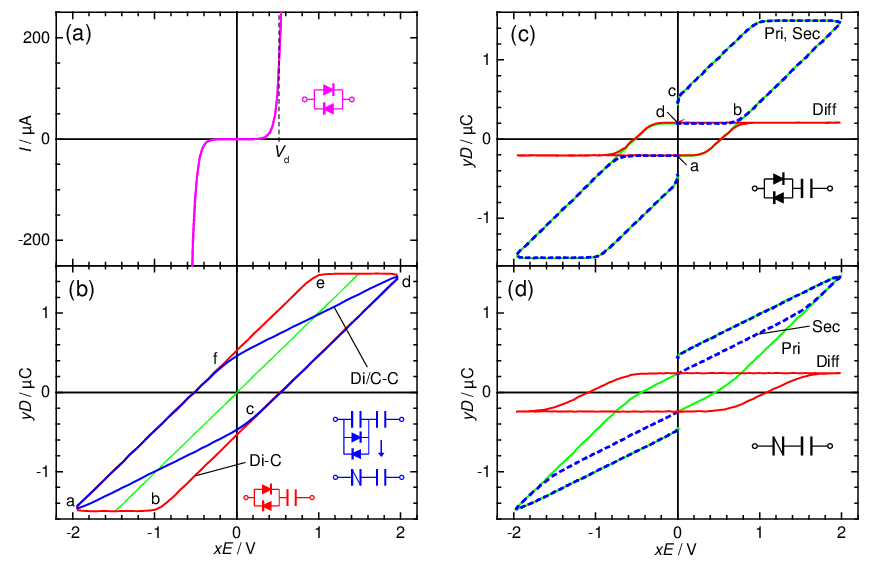}}
\caption{
(Color online) (a) An $I$--$V$ curve of the pair of diodes. 
(b) Continous $D$--$E$ loops of capacitors with the diodes. 
(c) DWM $D$--$E$ loops of the capacitor and diodes. 
(d) DWM $D$--$E$ loops of the capacitors with diodes, 
or a breakable capacitor in series.
}
\label{fig4}
\end{figure}

Figure \ref{fig4}(a) shows an $I$--$V$ curve of 
an antiparallel pair of silicon diodes, 
and Fig. \ref{fig4}(b) shows $D$--$E$ loops of a capacitor of $1 \rm{\mu F}$ 
with the diodes in series with continuous triangular waves of 20 Hz. 
Assuming that $I = 0$ at voltage of the diodes 
$|V\mathrm{_D}| < V\rm{_d}$ and $I = \pm \infty$ 
at $V\mathrm{_D} = \pm V\rm{_d}$, 
this case produces more simple apparent hysteresis loops 
than the case of a resistor shown later. 
Figure \ref{fig4}(b) also shows a $Q$--$V$ line of the capacitor only. 
This line indicates $V\rm{_C}$ of the capacitor at $Q = yD$, 
and a horizontal space between the line and the loop indicates $V\rm{_D}$. 
During the periods a-b and d-e on the loop labeled Di-C, 
$|V\mathrm{_D}| < V\rm{_d}$ and $D$ is constant. 
On the other hand, during b-c-d and e-f-a, 
the loop traces the lines $\pm V\rm{_d}$ apart from the line 
of the capacitor with the current $I = 160\ \rm{\mu A}$. 
(It takes 6.25 ms to change $xE$ by 1 V and $yD$ by $1 \rm{\mu C}$.) 
Hence a parallelogram-shaped fake loop is drawn. 
A circuit with an additional capacitor of the same capacity 
in parallel to the diodes draws another loop labeled Di/C-C. 
The capacitor with parallel diodes is considered 
as one breakable capacitor and simbolized as the inset. 
In this case, during the periods a-c and d-f, 
$I$ flows through the parallel capacitor 
and $yD$ changes as a series pair of the capacitors.

Figure \ref{fig4}(c) shows continuous (stable) DWM $D$--$E$ loops for Di-C. 
The Pri loop is drawn with the same beginning as the Diff loop, 
and almost coincides with the loop shown in Fig. \ref{fig4}(b) 
except for the beginning. 
The Pri loop traces a-b-c and the Sec loop traces d-b-c during $E > 0$, 
and the Diff loop traces as shown in the figure. 
The period c-d is an interval between the end of P and the start of U, 
namely 300 ms $-$ 25 ms. 
The charge $yD$ on the capacitor decreases during this period 
because of a little current through the diodes, 
and the apparent $P\rm{_r}$ decreases as the interval of the DWM increases. 

\begin{figure}
\centerline{\includegraphics[width=8.6cm]{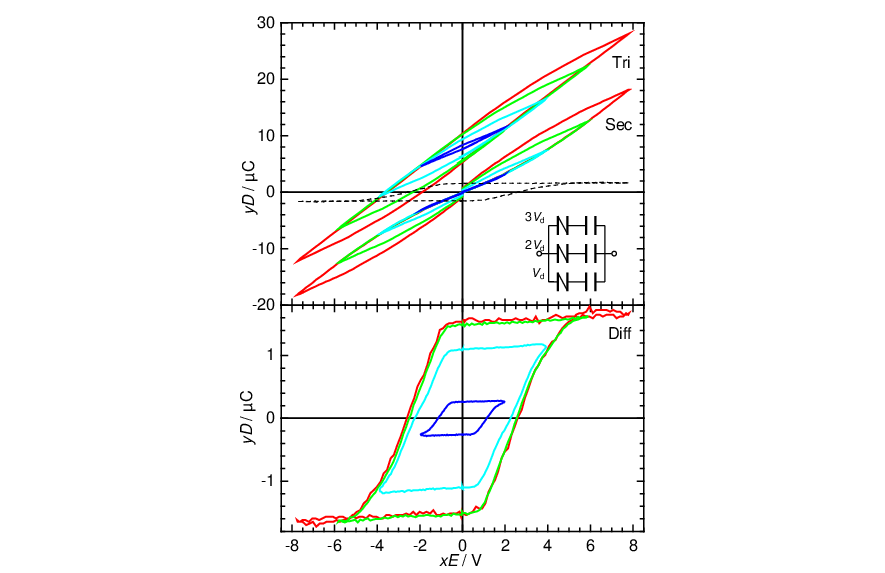}}
\caption{
(Color online) 
Continous triangular and DWM $D$--$E$ loops of a circuit model 
as the inset to imitate those of the MLCC.
}
\label{fig5}
\end{figure}

Figure \ref{fig4}(d) shows DWM $D$--$E$ loops for Di/C-C. 
In comparison with Fig. \ref{fig4}(c), 
the tops and the bottoms of Pri and Sec loops slant, 
and the apparent $E\rm{_c}$ becomes higher. 
These fake loops are partly similar to the loops of the MLCC. 
The $P\rm{_r}$ of the MLCC does not saturate with higher $V_0$, 
and a circuit in which these components with various $V\rm{_d}$ 
(adding diodes in series) are paralleled was measured. 
Figure \ref{fig5} shows the result. 
Although the apparent $P\rm{_r}$ saturates, 
similar $D$--$E$ loops to the MLCC are observed. 
Ferroelectric-like $D$--$E$ loops of the MLCC are considered 
as the apparent due to partly breakable capacitor,
which is, however, just one hypothesis.

\begin{figure}
\centerline{\includegraphics[width=8.6cm]{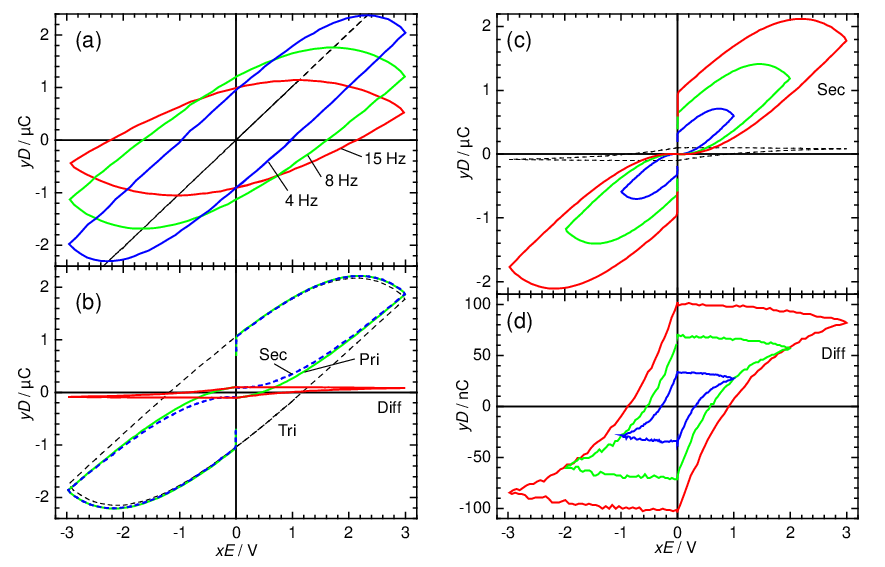}}
\caption{
(Color online) 
$D$--$E$ loops of a capacitor with a series resistor. 
(a) By continuous triangular waves with various frequencies. 
(b) By the DWM at 5 Hz with the interval of 150 ms. 
(c) Sec loops the same as shown in (b) with various amplitudes. 
(d) Enlarged Diff loops shown in (b).
}
\label{fig6}
\end{figure}

Figure \ref{fig6} shows $D$--$E$ loops of a capacitor 
($C = 1\ \mathrm{\mu F}$) with a series resistor ($R = 20\ \mathrm{k\Omega}$). 
Figure \ref{fig6}(a) shows the loops with continuous triangular waves 
of various frequencies. 
Although the slopes of the loops seem to decrease 
as frequency increases, 
these loops are considered to be horizontally inflated 
like the loops with diodes shown in Fig. \ref{fig4}(b), 
but the space namely $V\mathrm{_R} = IR$ changes all the time. 
Apparent $P\rm{_r}$ becomes maximum at the frequency of 
$1/2 \pi RC = 7.96\ \rm{Hz}$, 
which is calculated from the AC admittance. 

Figure \ref{fig6}(b) shows $D$--$E$ loops of the same circuit by the DWM. 
Here the frequency and the interval are set 5 Hz and 150 ms, 
respectively to increase the apparent $P\rm{_r}$. 
Because the charge on the capacitor continuously decreases 
through the resistor, 
this apparent $P\rm{_r}$ disappears when the interval is long enough. 
Figure \ref{fig6}(c) shows Sec loops with various amplitudes, 
and Fig. \ref{fig6}(d) shows enlarged Diff loops. 
Although Diff loops exhibit hysteresis, 
$D$ changes monotonically from $E = 0$ to maximum, 
and $D$ in the upper and the lower parts of the loops 
is not constant but increases as $|E|$ decreases. 
Moreover, the loops become simply large in the same shape 
as the amplitude increases. 
Sec loops also show unnatural characteristics as $D$--$E$ loops. 
The slope at the beginning of the loop becomes small or negative, 
which contradicts the permittivity. 
Sec loops are not meaningless, 
but few papers using the DWM show Sec loops.

Several papers\cite{FDE1,FDE2,FDE3} using the DWM show similar Diff loops 
as shown in Fig. \ref{fig6}(d). 
Although resistance in series might be poor electrodes on the samples, 
probably the materials are similar kinds 
which show a high apparent permittivity 
but it decreases at a low temperature  
because of highly temperature-dependent conductivity.\cite{CDR}

\begin{figure}
\centerline{\includegraphics[width=8.6cm]{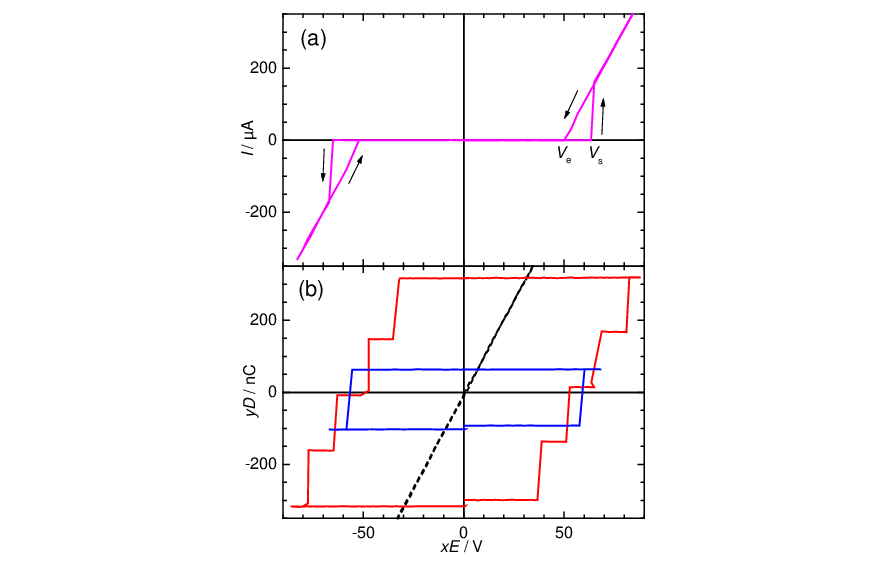}}
\caption{
(Color online) 
(a) An $I$--$V$ loop of a neon lamp with a resistor in series. 
(b) $D$--$E$ loops of a capacitor with this neon lamp in series 
with a single triangular wave.
}
\label{fig7}
\end{figure}

Figure \ref{fig7} shows another case of fake loops 
with a small neon lamp or bulb. 
Figure \ref{fig7}(a) shows an $I$--$V$ loop of a neon lamp 
with a resistor of 100 k$\Omega$ in series. 
The current does not flow at the lamp voltage 
$V\mathrm{_L} < V\mathrm{_s} = 65\ \mathrm{V}$ 
The lamp turns on and $I$ rapidly increases at
$V\mathrm{_L} = V\mathrm{_s}$ 
and the resistor limits $I$ at $V > V\mathrm{_s}$. 
When $V$ decreases, the lamp keeps on until 
$V\mathrm{_L} = V\mathrm{_e} = 50\ \mathrm{V} < V\mathrm{_s}$.
A mere gap in wiring can draw a similar $I$--$V$ curve 
when it sparks by a high voltage, but it is unstable. 
Figure \ref{fig7}(b) shows $D$--$E$ loops of a capacitor of 10 nF 
with this neon lamp in series by a single triangular wave. 
In this case, $D$ increases rapidly from 
$V\mathrm{_L} = V\mathrm{_s}$ to $V\mathrm{_L} = V\mathrm{_e}$,
and $D$ changes like steps. 
The steps repeatedly appear as the voltage becomes higher. 
When the DWM is used, Sec loops are almost zero, 
and Diff loops and Pri loops are almost the same 
as Tri loops shown in Fig. \ref{fig7}(b). 
This kind of apparent loops can appear because of bad wiring to a sample. 
Similar $D$--$E$ loops are shown in Ref. \citen{FDE4}.

\subsection{Measurement of $S$--$E$ loops with an EM}

\begin{figure}
\centerline{\includegraphics[width=8.6cm]{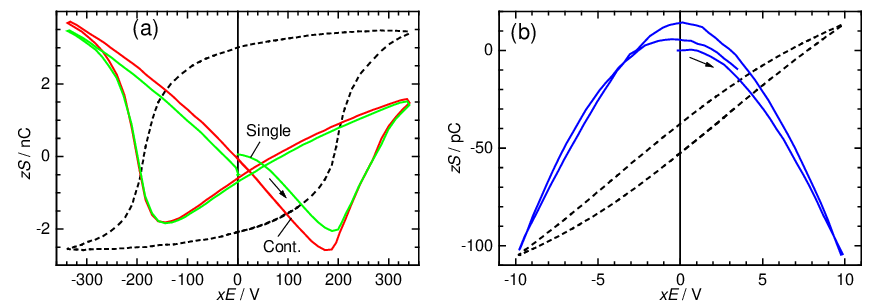}}
\caption{
(Color online) 
$S$--$E$ loops by an EM with triangular waves at 20 Hz. 
Broken curves are $D$--$E$ loops measured with the same condition, 
though their vertical scales are omitted. 
(a) PZT. (b) MLCC.
}
\label{fig8}
\end{figure}

Figure \ref{fig8}(a) shows $D$--$E$ and $S$--$E$ loops of the PZT sample 
in the same condition. 
The vertical axis of $S$--$E$ loops is written as $zS$ similar to $yD$. 
$S$ inverts across $E\mathrm{_c}$ and peculiar loops 
called butterfly loops are observed. 
Although $D$--$E$ loops can measured by one cycle of triangular waves 
or the DWM, $S$--$E$ loops with an EM show a little change 
at the beginning as ``Single'' in the figure. 
The reason is unknown, 
and $S$--$E$ loops later are measured by continuous triangular 
waves as ``Cont''. 
These $S$--$E$ loops seem to lean to right down, 
which indicates that an irreversible polarization remains in the sample. 
The irreversible polarization causes asymmetric 
$D$--$E$ and $S$--$E$ loops to $E$. 
Compared with an $S$--$E$ loop shown in Ref.\ \citen{PZT}, 
the maximum strain and the displacement are estimated to be 0.25\% 
and 500 nm, respectively, 
and the EM output is estimated to be 250 nm/nC.

Figure \ref{fig8}(b) shows $D$--$E$ and $S$--$E$ loops of the MLCC 
set together with the PZT. 
A parabolic $S$--$E$ loop appears even with a small voltage of 10 V, 
which means electrostriction proportional to $E^2$. 
$S$ becomes negative but the sample elongates along the $E$ 
because of the sample setup. 
Inversion of $S$ corresponding to polarization reversal is not seen, 
and $D$--$E$ loops of the MLCC sample are considered 
as fake ferroelectric loops.

\begin{figure}
\centerline{\includegraphics[width=8.6cm]{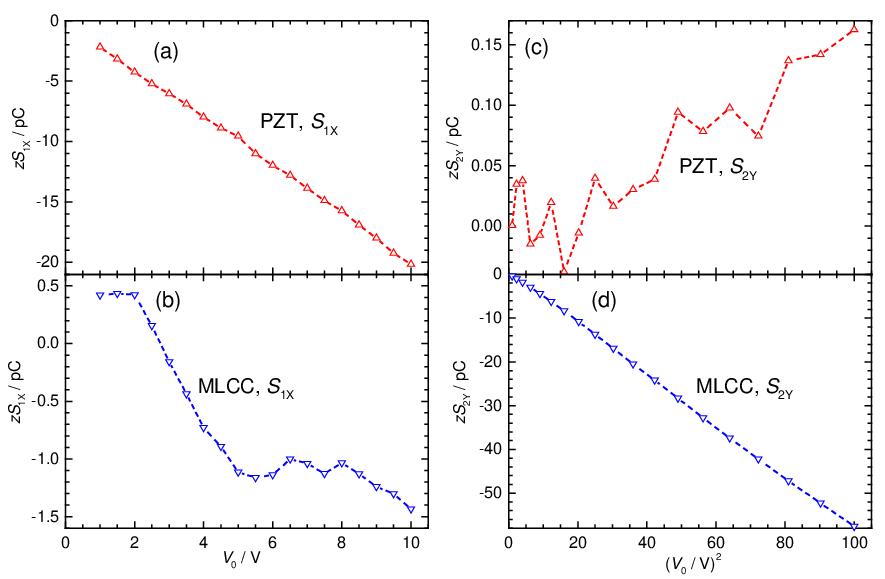}}
\caption{
(Color online) 
AC strain components of the PZT and MLCC samples 
with various AC voltages at 20 Hz. 
(a) $S\mathrm{_{1X}}$ of the PZT (b) $S\mathrm{_{1X}}$ of the MLCC. 
(c) $S\mathrm{_{2Y}}$ of the PZT. (d) $S\mathrm{_{2Y}}$ of the MLCC.
}
\label{fig9}
\end{figure}

Figure \ref{fig9} shows AC strain components of the PZT and MLCC samples 
with AC voltages $V_0$ up to 10 V at 20 Hz. 
In this case, $S\mathrm{_{1X}}$ of the PZT is negative because of 
the negative remanent polarization. 
$S\mathrm{_{1X}} \propto V_0$ is dominant for the PZT, 
while $S\mathrm{_{2Y}} \propto V_0^2$ is dominant for the MLCC, 
and values of them are comparable. 
$zS\mathrm{_{1X}}$ = 2 pC at $V_0$ = 1 V for the PZT, 
which means that the EM can sense the displacement of 0.5 nm
as the above estimation.

\begin{figure}
\centerline{\includegraphics[width=8.6cm]{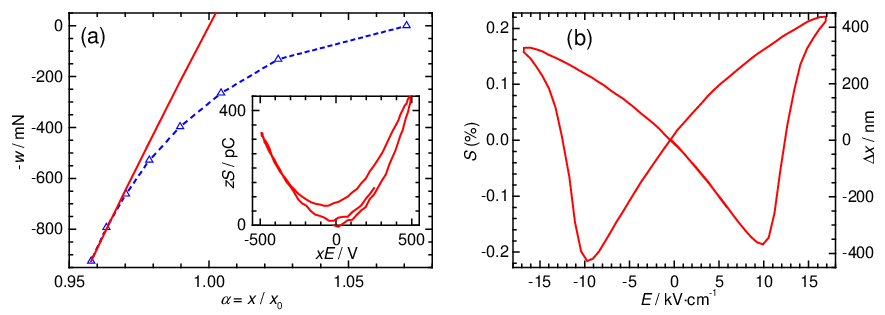}}
\caption{
(Color online) 
(a) The relationship between the air layer thickness and weight. 
The solid curve indicates rubber elasticity. 
(Inset) The EM signal due to attraction of the air layer capacitor. 
(b) An $S$--$E$ loop of the PZT sample at 50 Hz 
with vertical scales of strain and displacement.
}
\label{fig10}
\end{figure}

To calibrate $S$--$E$ loops by the EM, 
a comparison between two samples as shown in Fig. \ref{fig2}(a) has troubles 
such as an unstable relationship of the magnitudes of $zS$ of the samples. 
Therefore an air layer capacitor shown in Fig. \ref{fig2}(b) was used.
The inset in Fig. \ref{fig10}(a) shows the EM signal 
with applying AC voltage of $V_0 = 500\ \rm{V}$ 
at 50 Hz to the air layer capacitor as an $S$--$E$ loop. 
It shows a parabolic curve due to attraction proportional to $V^2$, 
and the height of the parabola was measured as 
$zS\mathrm{_A} = 2zS\mathrm{_{2Y}} = 330\ \rm{pC}$.
The maximum of the attraction corresponding to this signal 
was calculated as $F\rm{_A} = 5.4\ \rm{mN}$. 
Figure \ref{fig10}(a) shows the relationship between the air layer thickness 
$\alpha = x/x_0$ and weight $w$. 
The axes are chosen to correspond to rubber elasticity 
written as $F = p(\alpha - \alpha ^{-2})$. 
Results were not fit to the curve well at light $w$, 
probably because the surface of the rubber ring is not smooth 
and partially pressed. 
The solid curve was determined as $x_0 = 260\ \mu \rm{m}$ 
and $p = 7.0\ \rm{N}$. 
The thickness of the air layer with the sample and the vise 
was $x = 200\ \rm{\mu m}$, the pressure was $F\rm{_P} = -6.1\ \rm{N}$, 
and the elasticity at that point was 
$k = \mathrm{d}F/\mathrm{d}x = 140\ \rm{kN/m}$. 
The thickness change due to the attraction 
$\mathrm{\Delta} x\mathrm{_A}= F\mathrm{_A}/k = 38\ \rm{nm}$. 
The relationship between the EM output $zS$ and the displacement 
$\mathrm{\Delta} x$ was determined as 
$X = \mathrm{\Delta} x \mathrm{_A}/zS\mathrm{_A} = 110\ \rm{nm/nC}$. 
Figure \ref{fig10}(b) shows a calibrated $S$--$E$ loop of the PZT sample 
with this $X$. 
The strain obtained here is approximately 80\%
of that shown in Ref.\ \citen{PZT}.

\section{Conclusions}

It is demonstrated that apparent ferroelectric $D$--$E$ hysteresis loops 
can appear even with the DWM. 
Fake $D$--$E$ loops can be made with various circuit models, 
and it is recommended that users of the DWM try. 
Characteristics of the apparent loops appear in 
not only Diff loops but also Sec loops, 
and Sec loops by the DWM should be shown.
Apparent loops due to series resistance only are avoidable 
when the DWM interval is long enough.
A simple method to observe $S$--$E$ loops with an EM 
is developed as another method to confirm ferroelectricity. 
Although its calibration is not simple, 
$S$--$E$ loops can be qualitatively observed for samples 
of which $D$--$E$ loops are observed at room temperature 
with a little expansion. 
A commercial MLCC showing suspicious $D$--$E$ loops 
is revealed to be non-ferroelectric by $S$--$E$ loops.
The apparent ferroelectric-like $D$--$E$ loops of the MLCC 
are probably because of electrically breakable capacitors in it.


\end{document}